\documentclass{article}

\usepackage{graphicx}
\usepackage{arxiv}

\usepackage[utf8]{inputenc} 
\usepackage[T1]{fontenc}    
\usepackage{hyperref}       
\usepackage{url}            
\usepackage{booktabs}       
\usepackage{amsfonts}       
\usepackage{nicefrac}       
\usepackage{microtype}      
\usepackage{lipsum}		

\usepackage{doi}

\title{Minimal numerical ingredients describe chemical microswimmers' 3-D motion }

\author
{
Maximilian R. Bailey,\textit{$^{a,b}$} C. Miguel Barriuso Gutiérrez,\textit{$^{b}$} José Martín-Roca,\textit{$^{b,c}$}\\ Vincent Niggel,\textit{$^{a}$} Virginia Carrasco-Fadanelli,\textit{$^{d}$} Ivo Buttinoni,\textit{$^{d}$}\\ Ignacio Pagonabarraga,\textit{$^{e,f}$} Lucio Isa,\textit{$^{a}$} and Chantal Valeriani\textit{$^{b,g,\ast}$} 
\\
\small{$^{a}$Laboratory for Soft Materials and Interfaces, Department of Materials}\\
\small{ETH Z{\"u}rich, Z{\"u}rich, Switzerland}\\
\small{$^{b}$Departamento de Estructura de la Materia, Física Térmica y Electrónica,} \\
\small{Universidad Complutense de Madrid, Madrid, Spain}\\
\small{$^{c}$Departamento de Química Física, Facultad de Química,} \\
\small{Universidad Complutense de Madrid, Madrid, Spain}\\
\small{$^{d}$Department of Physics, Institute of Experimental Colloidal Physics,} \\
\small{Heinrich-Heine University, Düsseldorf, Germany}\\
\small{$^{e}$Departament de Física de la Matèria Condensada,} \\
\small{Universitat de Barcelona, Barcelona, Spain}\\
\small{$^{f}$Universitat de Barcelona Institute of Complex Systems (UBICS),} \\
\small{Universitat de Barcelona, Barcelona, Spain}\\
\small{$^{g}$GISC - Grupo Interdiplinar de Sistemas Complejos} \\
\small{Madrid, Spain}\\
\small{$^\ast$To whom correspondence should be addressed:} \\
\small{E-mail: cvaleriani@ucm.es}
}

\begin{document}
\maketitle

\begin{abstract}
The underlying mechanisms and physics of catalytic Janus microswimmers is highly complex, requiring details of the associated phoretic fields and the physiochemical properties of catalyst, particle, boundaries, and the fuel used.  Therefore, developing a minimal (and more general) model capable of capturing the overall dynamics of these autonomous particles is highly desirable. In the presented work, we demonstrate that a coarse-grained dissipative particle-hydrodynamics model is capable of describing the behaviour of various chemical microswimmer systems. Specifically, we show how a competing balance between hydrodynamic interactions experienced by a squirmer in the presence of a substrate, gravity, and mass and shape asymmetries can reproduce a range of dynamics seen in different experimental systems. We hope that our general model will inspire further synthetic work where various modes of swimmer motion can be encoded via shape and mass during fabrication, helping to realise the still outstanding goal of microswimmers capable of complex 3-D behaviour.
\end{abstract}

\section{Introduction}

In the last two decades, potential applications for directed transport at length scales where thermal fluctuations are important have prompted the development of a range of synthetic microswimmers, each with its intricacies \cite{Ramaswamy2010,Bechinger2016,Gompper2020}. Amongst the various synthetic active materials, Janus catalytic microswimmers remain one of the most popular due to their straightforward fabrication protocols, simple experimental set-ups, and good reproducibility between experiments \cite{Howse2007,Wang2020}. Typically, these are spherical particles asymmetrically modified with a catalytic material leading to the production of asymmetric local chemical gradients in the presence of a ``fuel'', causing propulsion via self-phoresis. \cite{Golestanian2005,Popescu2010,Dey2016, Bailey2021b}. Such microswimmers generally move in 2-D (xy) due to their density mismatch with the surrounding fluid and attractive interactions with the underlying substrate \cite{Uspal2015}. However, controlling the motion of microswimmers in 3-D is appealing from an applications perspective, and there is a growing body of work on active materials capable of motion in all dimensions \cite{Campbell2013,Campbell2017,Singh2018,Yasa2018,Lee2019,Brooks2018,Bailey2023,Bailey2021a,Carrasco2023a}. 

The rational design of chemical microswimmers displaying tailored motion in 3-D would greatly profit from models which can capture empirical observations \cite{Brooks2018}. However, their experimental simplicity masks a complex system of chemical and mass transfer relationships, the underlying mechanisms and physics of which are still the subject of ongoing debate \cite{lyu2021,Dominguez2022}. It is generally accepted that a more complete description of such “chemically active colloids” requires the full solution of their phoretic fields, including details of the colloids, the substrate, and the solution composition \cite{Popescu2018a,Katuri2021,Torrenegra_2022}. Unfortunately, such detailed descriptions call for a high level of technical expertise, are system-specific due to phoretic mobility parameters, have only been solved for spherical and ellipsoidal structures, and do not easily allow the inclusion of thermal fluctuations, which are critical when describing the dynamics of micron-scale objects. 

To avoid the consideration of chemical phoretic fields and only account for hydrodynamic flows, the “squirmer" model is frequently invoked \cite{Zottl2023}. This model was first proposed for microorganisms such as \textit{Paramecia} and \textit{Volvox} \cite{Lighthill1952,Blake1971}, and is now used as a generic description for various active systems. Squirmers swim due to a self-generated, usually stationary \cite{Magar_2003} and axi-symmetric velocity field which is typically evaluated as tangential across its surface. Considered as a spherical rigid body, the squirmer can be defined using two modes describing its swimming velocity and force-dipole (B\textsubscript{1} and B\textsubscript{2} respectively \cite{Theers2016}). Promisingly, it has been shown that a bottom-up model of Janus self-diffusiophoretic microswimmers - accounting for the unique interaction potentials of the different chemical species with the separate hemispheres of a Janus particle - will produce flow fields characteristic of spherical squirmers \cite{Yang2014}. Recent experimental studies investigating tracer flows around chemical microswimmers \cite{Campbell2019} and their directed motion under flow (“rheotaxis") \cite{Katuri2018a,Sharan2022b} have also indicated that the flow fields generated by such active agents are characteristic of squirmer-type systems. Therefore, “coarse-graining" Janus chemical microswimmers as squirmers provides a viable approach to model their behaviour despite neglecting the (albeit important) contributions arising from chemical fields \cite{Katuri2021,Torrenegra_2022} and an asymmetric surface \cite{Yang2014}.

A number of numerical strategies have been implemented to simulate the dynamics of squirmers, amongst which a multi-particle-collision dynamics (MPCD) description of the solvent is perhaps the most frequently invoked due to its ability to include thermal fluctuations at a lower computational cost than e.g. the Lattice Boltzmann method \cite{Llopis_2010,Pagonabarraga_2013,Alarcon_2019,Scagliarini_2022,Theers2016,Yang2014,Malevanets1999,Malevanets2000,Downton2009,Zottl2018a}. Like MPCD, dissipative particle dynamics (DPD) \cite{Groot1997} coarse-grains the solvent as point-like fluid particles (packets of fluid molecules), and thus inherently includes the effects of thermal noise while solving the Navier-Stokes equations \cite{Hoogerbrugge1992}. By utilising a particle-based approach to model the solvent, the computational cost is significantly reduced, allowing the simulation of hydrodynamic interactions in systems where advection dominates over diffusion (high Péclet number, Pe) \cite{Koelman1993}. In DPD, the solvent particles themselves act as local thermostats due to their competing stochastic and dissipative pair-wise terms, conserving momentum and thus accounting for hydrodynamics and thermal fluctuations \cite{Espanol1995}. Additionally, DPD makes use of softer inter-particle potentials, allowing greater timesteps compared to MPCD, thereby enabling simulations of larger systems at longer timescales \cite{Soddemann2003,Lugli2012}. Therefore, DPD emerges as a suitable candidate to model microswimmers as squirmers, as it can handle hydrodynamics in high Pe systems (where directed motion dominates over diffusion) with (relatively) large time-steps, while properly dealing with thermal fluctuations. For these reasons, a DPD framework capturing the hydrodynamic interactions of microswimmers in the presence of confining boundaries - by applying tangential solvent forces around the swimmer - was recently developed by some of the authors \cite{Barriuso2022}. As DPD is already implemented in the open-access molecular dynamics program Large-scale Atomic/Molecular Massively Parallel Simulator (LAMMPS) \cite{Plimpton1993}, this provides the opportunity to exploit a range of in-built functions to extend previous studies, with the goal of mimicking the behaviour of chemical microswimmers above a substrate.

Here, we further develop this model to consider the influence of mass and shape asymmetries on the dynamics of spherical microswimmers in the presence of a bounding substrate, mirroring common experimental conditions for chemical active colloids \cite{TenHagen2014,Simmchen2016,Das2020_A, Barriuso2022}. The strength of our approach lies in its modularity, which allows the simple inclusion of these asymmetries that may have otherwise presented significant complications when considering other numerical schemes. We find that the interplay between hydrodynamic interactions, gravity, bottom-heaviness, and shape is sufficient to qualitatively capture the 2- and 3-D physics of a range of catalytic (and photo-catalytic) Janus microswimmer systems \cite{Bailey2021a,Bailey2023,Carrasco2023a}, i.e. while neglecting contributions from chemical \cite{Das2020_A} and light \cite{Singh2018b,Uspal2019} gradients. Specifically, there is a competition between the hydrodynamic attraction to the substrate that grows with the swimming speed and the active force required to overcome gravity, which determines the ability of the microswimmer to enter the bulk. This balance can be furthermore adjusted by introducing mass or shape asymmetry to the particles. Our coarse-grained approach thus opens the door to the informed design of chemical microswimmers whose dynamics can be encoded via shape or mass during fabrication, helping to realise the still outstanding goal of active colloids capable of truly 3-D dynamics.
 
\section{Numerical method}

\begin{figure}
    \centering
    \includegraphics[width=0.5\linewidth]{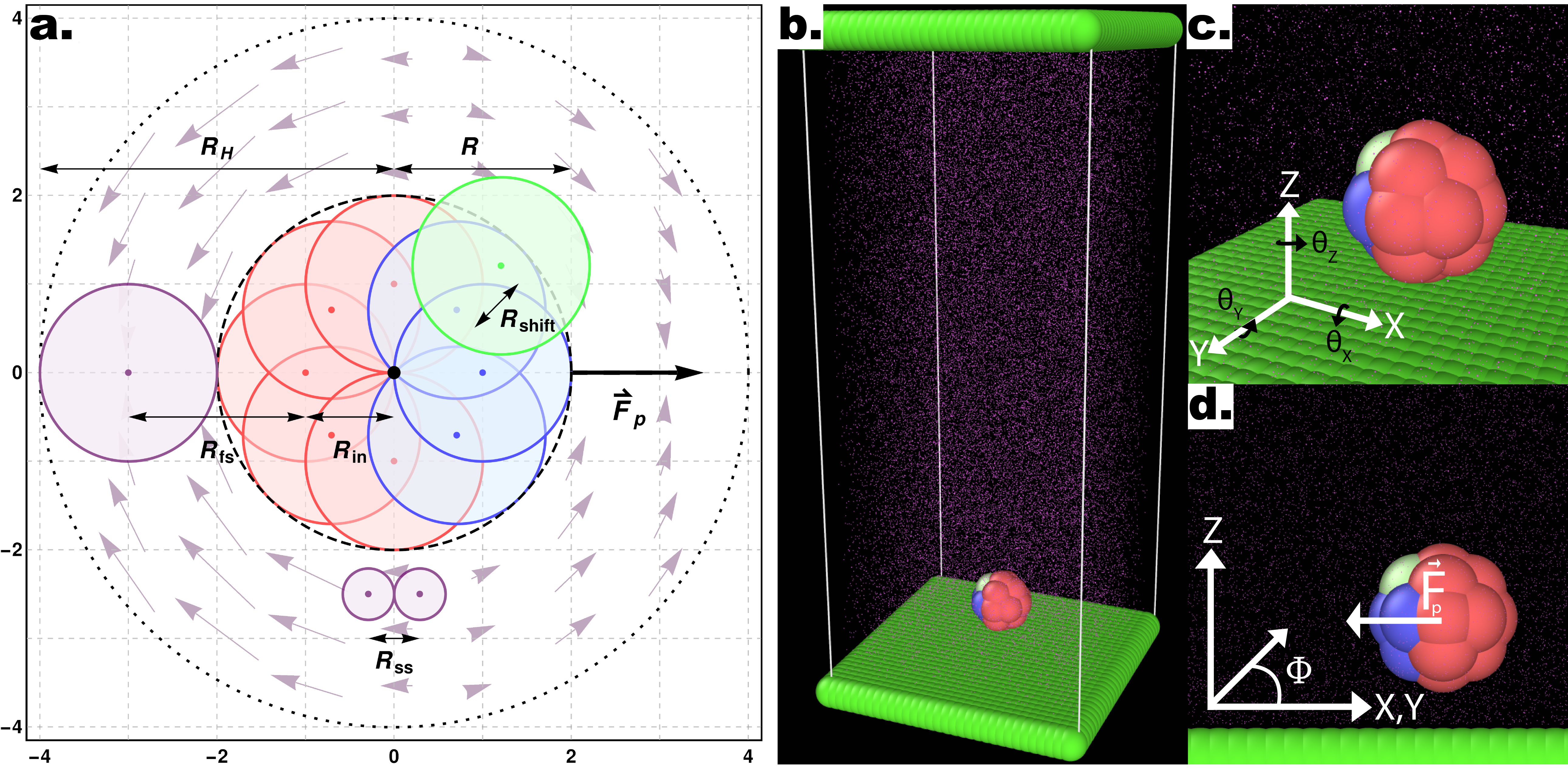}
 \caption{a) Schematic representation of a 2D section of the colloid using a raspberry model with cap-front (CF) mass imbalance and shape asymmetry. The colloid is built from 18 DPD \textit{filler} particles (red and blue), distributed on the surface of a sphere radius $R_{in} = 1$, each with a DPD cut-off for interactions with other DPD particles (e.g. solvent particles), resulting in an overall effective microswimmer radius of $R = 2$. Mass imbalance is introduced by increasing the mass of the particles constituting the cap (blue). Shape asymmetry is implemented by introducing an additional particle $P_{shape}$ (green) displaced a distance $R_{shift}$ from an off-axis filler particle. Here, the equatorial 2D slice containing the shape asymmetry is shown, thus only depicting the 8 filler particles within this section. Solvent particles (purple) interact with the colloid particles with a DPD cutoff of $R_{fs}=2$ while between them we set $R_{ss}=0.58$ to achieve a lower Reynolds number \cite{Groot1997}. The region in which the hydrodynamic propulsion takes place is the spherical shell between the outer surface of the microswimmer (dashed black circle) and $R_H = 4$ (dotted black circle), solvent particles within this region are subjected to a force field (light purple arrows), consistent with a pusher-type squirmer, emanating from the center of mass of the colloid and fixed with its internal frame of reference. Then, for each solvent particle in this region an equal and opposite reaction force is applied to its nearest colloid particle, resulting in a net self-propulsion force $\vec{F}_p$. b-d) Snapshot of the full simulation box (b.), as well as the definition of the angles $\theta_{x,y,z}$ (c.) and the azimuthal angle $\Phi$ (d.) Graphics presented here and elsewhere were generated in part using the visualisation software Ovito \cite{Stukowski2010}.}
  \label{fig:Fig_Schematic}
\end{figure}

Following the approach outlined in \cite{Barriuso2022}, we use an in-house extension of the open source LAMMPS package \cite{Plimpton1993} to simulate the motion of squirmer-like microswimmers, modelled as raspberry-type structures (see Figure \ref{fig:Fig_Schematic}). The thermal energy of the solvent particles $k_BT$ is set to $1$, the solvent density $\rho = 5.9$, and the dissipative force $\gamma = 1$. The simulation length-scales are as described in Figure \ref{fig:Fig_Schematic} a., normalised with respect to the radius of 1 DPD “filler" particle, while the masses of DPD particles are set to $1$ unless introducing mass asymmetry (see below). The simulation time-scale $\tau$ is set according to these parameters, and numerically integrated at $0.01 \tau$ time-steps. When present, substrates are modelled as overlapping DPD particles, properly aligned along the x- and y-axes to ensure a flat repulsive potential from the DPD conservative force $F_c = 100$. The microswimmers consist of 1 central “thruster" particle generating the flow fields which are experienced by the solvent particles within a hydrodynamic radius $R_h$ of the thruster, which in turn apply reaction forces to the “filler" particles (18-20 DPD particles, depending on whether or not shape asymmetry is introduced to the microswimmer, see Figure \ref{fig:Fig_Schematic} a.), resulting in the net propulsion force of the microswimmer. The mass of specific filler particles can be adjusted, allowing us to introduce mass asymmetry, and thus model microswimmers equipped with heavy "caps" (e.g. Pt-coated catalytic microparticles). We define mass asymmetry as $m_{asymm} = \frac{m_{hemisphere}+m_{cap}}{m_{hemisphere}}$ (where $m_{hemisphere}$ and $m_{cap}$ are the mass of a particle hemisphere and the heavy cap respectively), noting that the motion can thus be defined with the heavier cap at the back (CB) or at the front (CF) of the swimmer with respect to the swimming direction. In all cases, we use a squirmer active stress parameter $\beta = B_2/B_1 = -2.5$, i.e. a weak pusher, based on qualitative matching of simulated trajectories to the behaviour described in \cite{Carrasco2023a}. We note that this squirmer parameter corresponds to that experimentally determined by Campbell et al. \cite{Campbell2019} for a similar Pt catalytic system as that studied in \cite{Carrasco2023a}, which provides further evidence for the suitability of our selection. Solvent conditions are selected to ensure that all microswimmers with radius $R$ studied remain in the low Reynolds (Re) number regime ($Re = \frac{v_p\cdot R}{\nu} < 0.2$ \cite{Padding2006}, also see Supporting Information - Figure S1), where $v_p$ is the particle speed, and specifically by controlling for the kinematic viscosity $\nu$ via the solvent parameters \cite{Groot1997}. To map different simulations to experiments, we modulate the ratio of swimming velocity $V_{swim}$ to sedimentation velocity $V_{gravity}$. A minimum of 2000 time steps are used for equilibrating solvent conditions in all simulations, as were periodic boundary conditions. The number of solvent DPD particles depends on the simulation dimensions, ranging from 47002 for a (20x20x20, xyz) box, to 57197 solvent DPD particles for a (16x16x40) box.

\section{Results \& Discussion}
\subsection{Role of mass asymmetry and hydrodynamic interactions in microswimmer motion}
\begin{figure}[h!]
\centering
\includegraphics[width=0.75\linewidth]{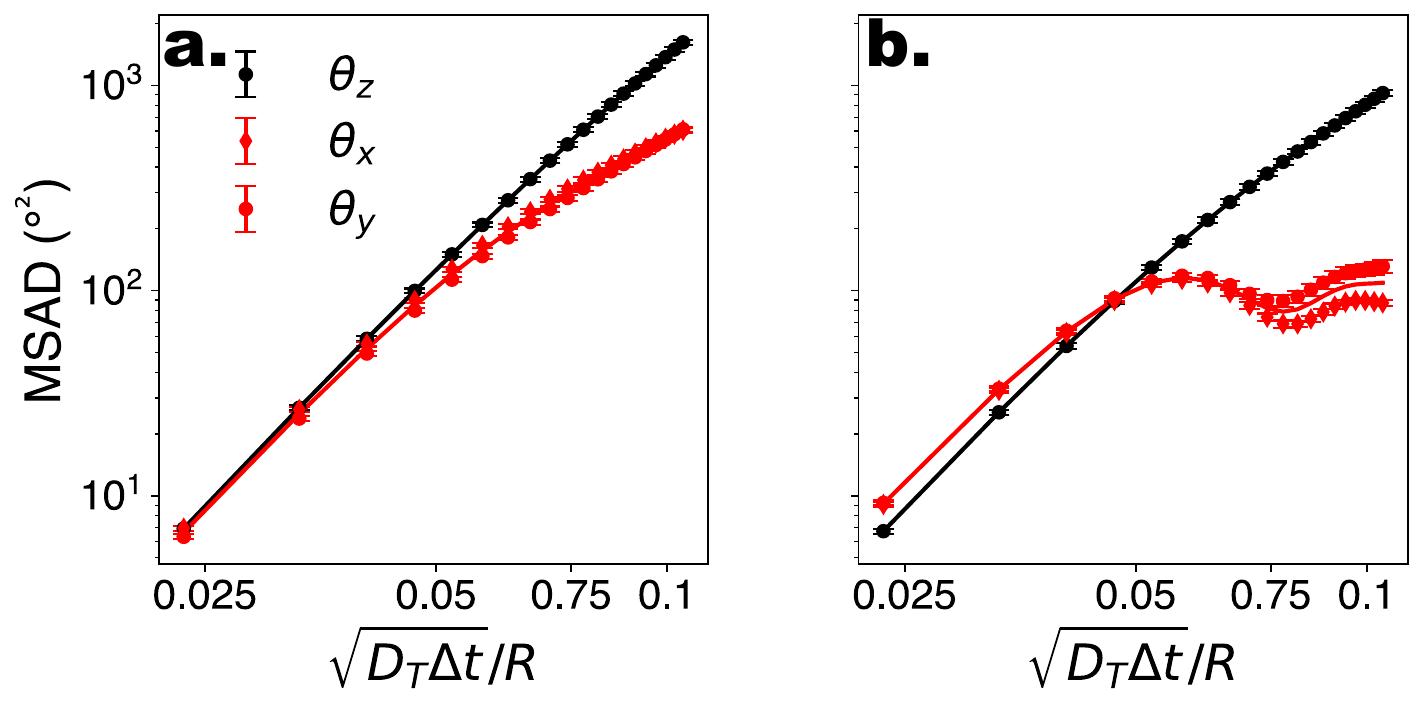}
  \caption{Rotational dynamics of CB microswimmers as a function of scaled time, simulated to reproduce the properties of the chemical active colloids studied in \cite{Niggel2023}. a) Mean-squared-angular-displacement (MSAD) of the passive colloid (no activity). b) Introducing activity to the microswimmer dampens its rotational diffusivity, particularly about the x- and y- axes, which saturate due to the hydrodynamic coupling to the substrate. We note the scaling of the MSAD with time here is with $\sqrt{\Delta t}$, and is therefore quadratic. Error bars depict the standard error of the mean from 2401 frames (sub-sampled from 50000 simulation steps).}
  \label{fig:Fig1}
\end{figure}

To begin, we consider colloids characterised by hydrodynamic interactions with a surface and bottom-heaviness, and study their dynamics with or without activity. Specifically, we reproduce the 'classic' Pt-SiO\textsubscript{2} Janus chemical microswimmers to determine whether our DPD raspberry particle model captures their dynamics. We investigate the chemical microswimmers studied by Niggel et al. \cite{Niggel2023}, as the 3-D rotations of the Janus particles with fluorescent surface asperities can be tracked via correlation-based image analysis. To reproduce their experimental findings, we simulate the microswimmers with CB motion and $m_{asymm} = 1.081$. Fitting the short-time mean-square-displacement (MSD) of our experimental particles ($MSD = 4\cdot D_T\cdot\Delta t + v_p^2\cdot\Delta t^2$, while acknowledging its shortcomings \cite{Bailey2022b}), we obtain a Péclet number $Pe = \frac{v_p\cdot R}{D_T} \sim 300$ (where $v_p, D_T$ are the fitted swimming speed and translational diffusion coefficient from the MSD, respectively, and $R$ is the particle radius), which we set to $\sim 100$ in simulations of our active microswimmers to ensure that they remain in the low Reynolds regime as described above. Following our experimental findings, we also set the gravitational force such that $V_{gravity} \sim V_{swim}$, for $V_{swim} = <v_0>$, where $v_0$ is the microswimmer's instantaneous velocity over time. As we have access to the structural coordinates of our DPD raspberries, we are able to follow their rotational dynamics via singular value decomposition (SVD) and calculate the cumulative mean-squared-angular-displacement (MSAD) (see Figure \ref{fig:Fig1}, \cite{Ilhan2020}). By doing so, we are able to reproduce the findings presented in \cite{Niggel2023}. Specifically, we find that for passive particles, the presence of a heavy cap and a substrate reduces the rotation “out-of-plane" ($\theta_{x,y}$) compared to in plane ($\theta_z$) rotations (see Figure \ref{fig:Fig_Schematic} c.) due to bottom-heaviness (see Figure \ref{fig:Fig1} a., \cite{Rashidi2020} ), while the introduction of activity saturates the out-of-plane rotation at much lower values (Figure \ref{fig:Fig1} b.). We attribute the confinement of possible rotations to the hydrodynamic attraction between the microswimmer and the substrate due to its generated flow fields \cite{Spagnolie2012}, mirrored by later theoretical investigations into the role of the produced phoretic fields \cite{Uspal2015}. Therefore, we see that the rotational dynamics of chemical microswimmers near confining boundaries are qualitatively well captured when only considering hydrodynamic interactions (and bottom-heaviness). 
 
\begin{figure}
\centering
  \includegraphics[width=0.95\linewidth]{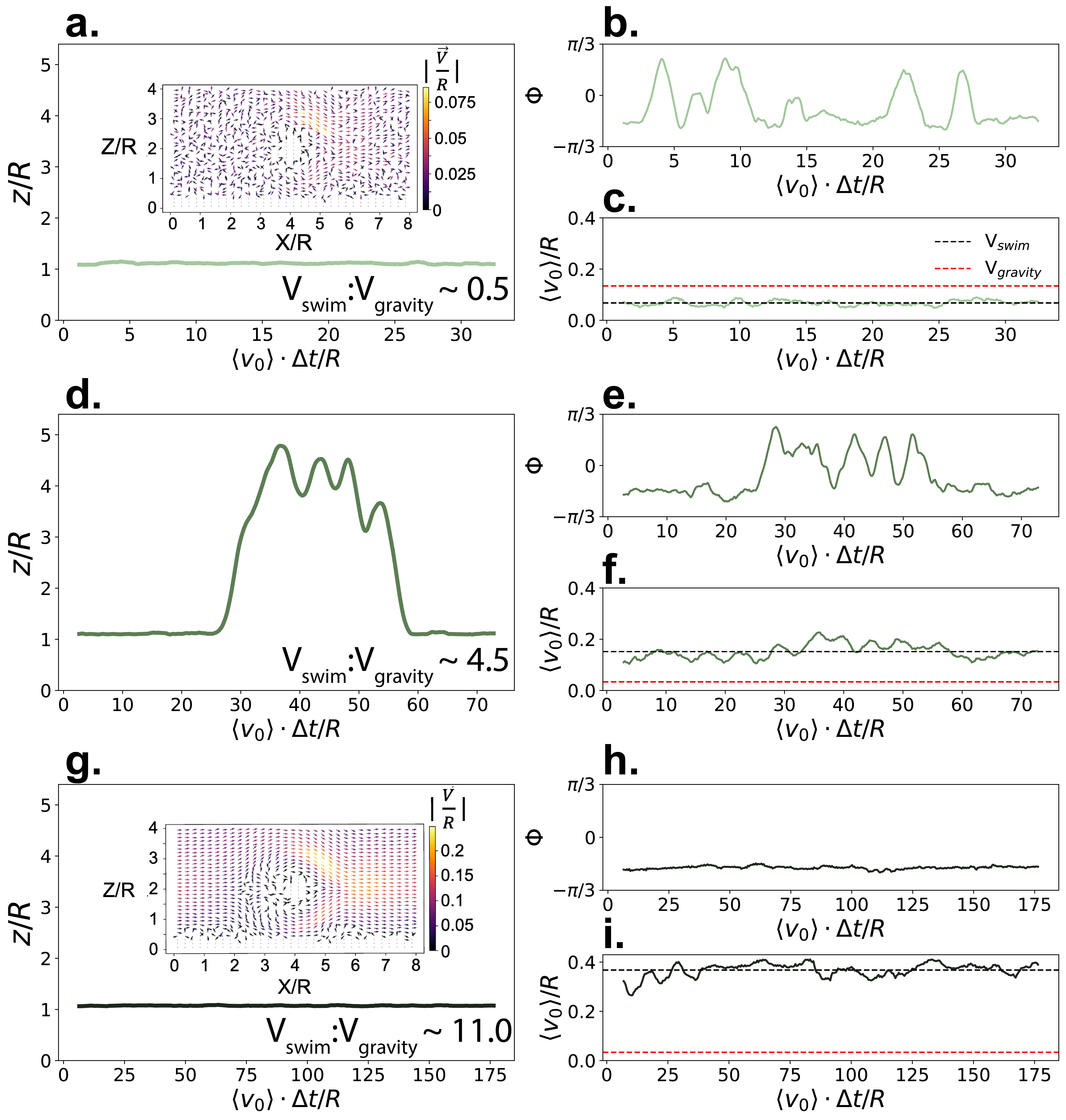}
  \caption{Dynamics of CB microswimmers, simulated to reproduce the properties of the chemical active colloids studied in \cite{Carrasco2023a}. $z/R$ refers to the height that a microswimmer reaches above the substrate with respect to its radius $R$ (a., d., and g.). $\Phi$ is the azimuthal angle of the microswimmer (b, e, and h) as defined in Figure \ref{fig:Fig_Schematic}, and $v_0$ is the time-series of the microswimmer's instantaneous velocity (c., f., and i., with $V_{swim} = <v_0>/R$). The insets in a. and g. show the solvent velocity flow fields around the (fixed) microswimmer at the different self-propulsion forces. Here, we progressively increase the swimming to sedimentation velocity ($V_{swim}:V_{gravity}$) from top to bottom to qualitatively map onto the different systems studied by Carrasco-Fadanelli and Buttinoni. The first two cases (a-c, d-f) are in good agreement with the findings presented in \cite{Carrasco2023a}, while the final case (g-i) recaptures the typical dynamics observed for chemical active colloids, where phoretic particles are essentially confined to 2-D (also see Figure \ref{fig:Fig1} b. ).
  }
  \label{fig:Fig2}
\end{figure}

We then map our squirmer-inspired DPD model to the experimental findings presented by Carrasco-Fadanelli and Buttinoni \cite{Carrasco2023a}. Specifically, we adjust $V_{swim}$ :$V_{gravity}$ to reproduce the different conditions studied in their work, while using the same value for mass asymmetry of their Janus polystyrene spheres coated with a Pt thin film (2.8 $\mu$m with a coating of 4 nm, $m_{asymm} = 1.22$) with CB motion. When $V_{swim} < V_{gravity}$, the gravitational torque from the heavier cap aligns the particle's internal orientation axis away from the substrate (see Figure \ref{fig:Fig2} b.), however the particle is unable to leave the substrate due to the force of gravity (see Figure \ref{fig:Fig2} a,c). As $V_{swim}$ is increased and becomes larger than $V_{gravity}$, the particle is able to leave the substrate when its internal orientation axis is directed upwards (see Figure \ref{fig:Fig2} d-f, and Supporting Information Figure S2), assisted by its bottom-heaviness \cite{Das2020_A}. Notably, we find that we can reproduce the structure of the microswimmer dynamics previously seen experimentally \cite{Bailey2021a,Bailey2023} (see Supporting Information Figure S3), without requiring the previously hypothesised self-shadowing effects \cite{Singh2018b,Uspal2019}. The two scenarios qualitatively capture the behaviour described in \cite{Carrasco2023a}, where the gravitational torque due to the mass asymmetry introduced by the denser Pt cap favours an anti-parallel orientation of the microswimmer with gravity, competing with hydrodynamic interactions and thermal fluctuations. Interestingly, by further increasing $V_{swim}$, the simulated microswimmer is once again confined to 2-D motion in the xy plane (see Figure \ref{fig:Fig2} g-i). However, here the confinement arises due to the internal orientation of the microswimmer, rather than its inability to overcome the applied gravitational force. For fast swimmers, the flow fields created by the microswimmer lead to a strong hydrodynamic attraction to the substrate, effectively “trapping" the particle in 2-D (compare insets in Figures \ref{fig:Fig2} a. and g.)  . This is similar to the “sliding-state" behaviour proposed by Uspal and coworkers as a result of hydrodynamic flows from self-generated phoretic gradients \cite{Uspal2015}, and mirrors the experimental observations of many chemical microswimmer systems, including that discussed in Figure \ref{fig:Fig1}. 

To summarise, we find that hydrodynamic interactions and gravitational forces on a mass-asymmetric spherical microswimmer are sufficient to qualitatively capture the physics of chemical microswimmers above a substrate under different conditions. We also confirm the importance of gravity on both the translational and rotational dynamics of active colloids suspended just above a planar surface.

\subsection{Influence of shape asymmetry on microswimmer 3-D dynamics}

\begin{figure}
\centering
  \includegraphics[width=0.95\linewidth]{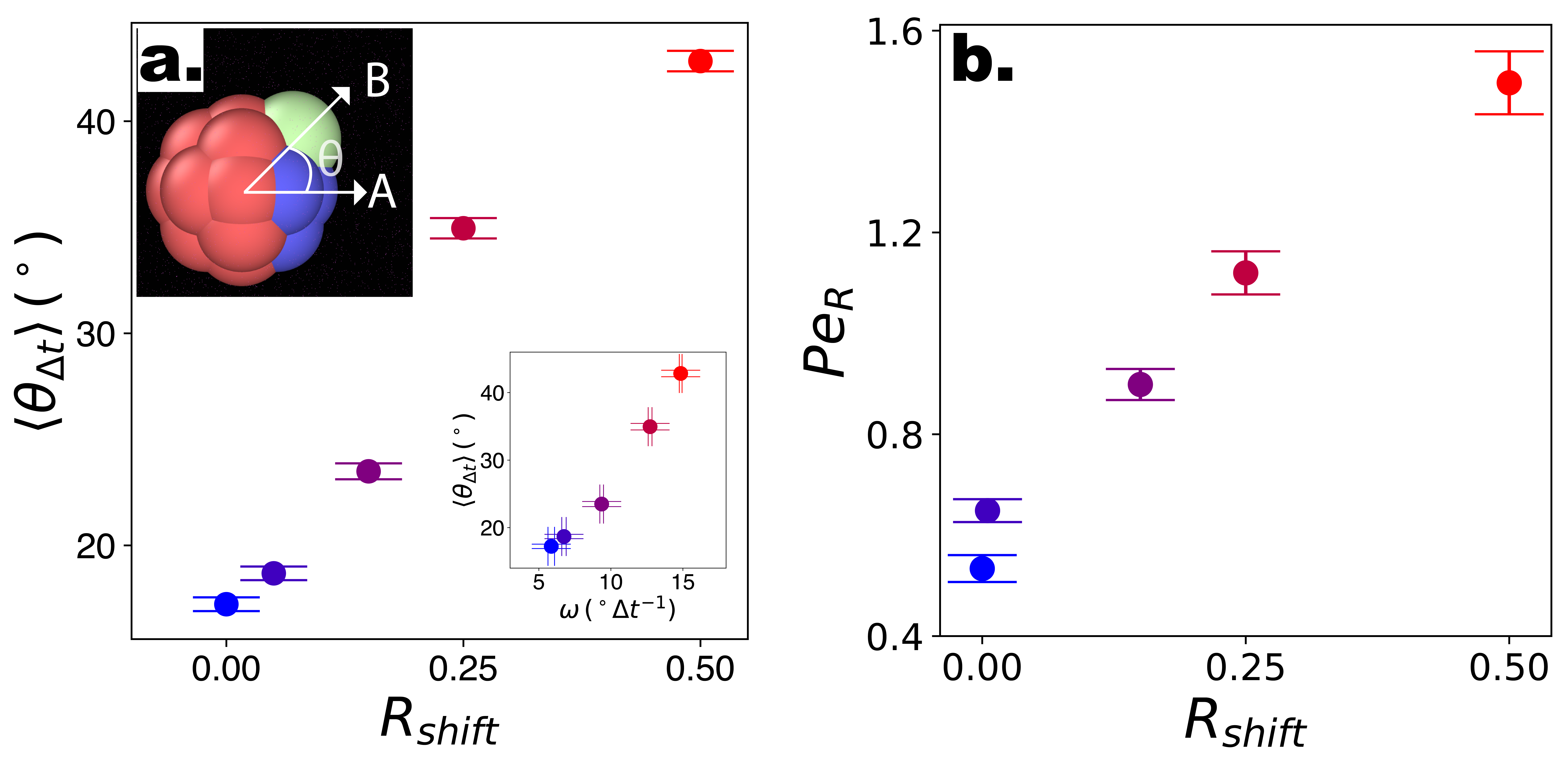}
  \caption{Swimming dynamics of CF microswimmers with respect to their cap orientation, in the absence gravity and bounding substrates, as a function of introduced shape asymmetry $R_{shift}$ (colour coded from blue to red with increasing $R_{shift}$). a) Increasing divergence $\langle\theta\rangle$ in the internal body axis and the observed swimming direction is measured with increasing $R_{shift}$. We note that the values of $\langle\theta\rangle$ are a function of the time-step evaluated (here, $\sqrt{D_{T}\Delta t}/R \sim 0.13$). Inset: linear growth in $\langle\theta\rangle$ with $\omega$ values fitted from the microswimmers' MSAD (see Supporting Information, Figure S6). Graphical inset: angle $\theta$ between the swimming direction and internal body axis, as defined in the text. The particle depicted has $R_{shift} = 0.5$. b) Rotational Pèclet number as a function of  $_{Rshift}$ for $\Delta t = 1$. Error bars either indicate the standard error of the mean, or for the fits of $\omega$ and $D_R$ are obtained from the co-variance matrix, accounting for error propagation where relevant, from 1001 frames (sub-sampled from 50000 simulation steps).}
  \label{fig:Fig3}
\end{figure}

Motivated by the ability of the reported DPD squirmer-like microswimmers to qualitatively reproduce the behaviour of different chemical microswimmers, we then investigate strategies to promote the “lift-off" of particles from the substrate, and thus observe 3-D motion. Recently, photocatalytic microswimmers synthesised using functional nanoparticles via “Toposelective Nanoparticle Attachment" were shown to display quasi 3-D behaviour \cite{Bailey2021a,Bailey2023}, characterised by 2-D motion interdispersed with “rollercoaster"-like looping behaviour. This happens in spite of their high speeds with respect to their sedimentation speed, putting them in a regime closer to the third case highlighted in Figure \ref{fig:Fig2} (g-i), as well as their CF motion which should in fact promote orientation into the wall due to top-heaviness. We note the use of surface-bound functional nanoparticles incorporates rough asperities to the microswimmers, which in turn introduce asymmetry in the drag acting on the surface of the microswimmers. In fact, explicitly considering non-axisymmetric flow fields in spherical squirmers was very recently shown to induce body rotations and thus complex patterns of motion \cite{Burada2022}. To investigate the potential effect of such shape-asymmetries on the motion of microswimmers, we simulate the motion of our DPD raspberries with introduced shape asymmetry in the absence of gravity or bounding substrates, this time with CF swimming (see Figure \ref{fig:Fig3}). Specifically, we introduce weightless “shape" particles ($P_{shape}$) along the axis of the existing “filler" particles, shifted by different distances (normalised by the DPD particle radius - $R_{shift}$) to control the extent of its protrusion (see Figure \ref{fig:Fig_Schematic} a.). The filler particle axis along which the shape particle is introduced can be changed to modify the extent of asymmetry, i.e. with respect to the swimming direction (See Supporting Information SI S5). We focus on the case where $P_{shape}$ is introduced so as to maximise the shape-asymmetry possible using this approach (see Figure \ref{fig:Fig_Schematic} a., Supporting Information Figure S5, “1 Edge" case). 

To quantify the effect that increasing $R_{shift}$ has on the dynamics of our bulk microswimmers, we calculate the angle, $\theta$, between the vectors describing the particle's internal orientation axis, $\textbf{A}$, (governing the direction of the self-propulsion force, see Figure \ref{fig:Fig_Schematic} a.), and the particles displacement vector, $\textbf{B}$, ($\theta = arccos(\frac{\textbf{A}\cdot \textbf{B}}{|A||B|})$), where $\textbf{A}$ is defined by 3 particles along the microswimmers body axis at time $t$ and $\textbf{B} = [x_{t+\Delta t} - x_{t}, y_{t+\Delta t} - y_{t}, z_{t+\Delta t} - z_{t}$). We note that the value of $\langle\theta\rangle$ is dependent on the $\Delta t$ over which the displacement vector is evaluated. We find a positive correlation between $R_{shift}$ and $\langle\theta\rangle$ (coefficient of determination$ = 0.953$, see Figure \ref{fig:Fig3} a.), which we attribute to the increasing torque experienced by the microswimmer due to the solvent forces applied to $P_{shape}$ at the distance $R_{shift}$. We hypothesise that the non-zero value for $\langle\theta\rangle$ at $R_{shift} = 0$ arises due to the presence of thermal fluctuations from the solvent and their effect on microswimmer motion (see Supporting Information Figure S4 for further discussion). The source of the growing divergence between the internal orientation of the particles and their swimming velocity can also be observed in the MSAD of the microswimmers as the trends become increasingly ballistic with larger $R_{shift}$ (see Supporting Information S6). Notably, the fitted angular velocity $\omega$ (ballistic component of the MSAD, $MSAD = 2\cdot D_R\Delta t + \omega^2\cdot\Delta t^2$, where $D_R$ is the rotational diffusion coefficient) grows with shape asymmetry, and we determine a strong linear relationship between $\omega$ and $\langle\theta\rangle$ (see Figure \ref{fig:Fig3} a., bottom inset - coefficient of determination$ = 0.981$). We furthermore note a linear growth of the “rotational" Péclet ($Pe_R = \frac{\omega\cdot\Delta t}{\sqrt{D_r\cdot\Delta t}}$) with $R_{shift}$ (see Figure \ref{fig:Fig3} b., coefficient of determination$ = 0.982$), demonstrating the increasingly deterministic orientational motion of the microswimmer as shape asymmetry becomes more pronounced. Finally, we note that introducing $P_{shape}$ along axes resulting in a reduced shape asymmetry with respect to $\textbf{A}$ have a negligible effect on $\theta$, indicating the importance of appropriate shape selection (see Supporting Information Figures S5, S7, and S8). 

\begin{figure}[h]
\centering
  \includegraphics[width=0.95\linewidth]{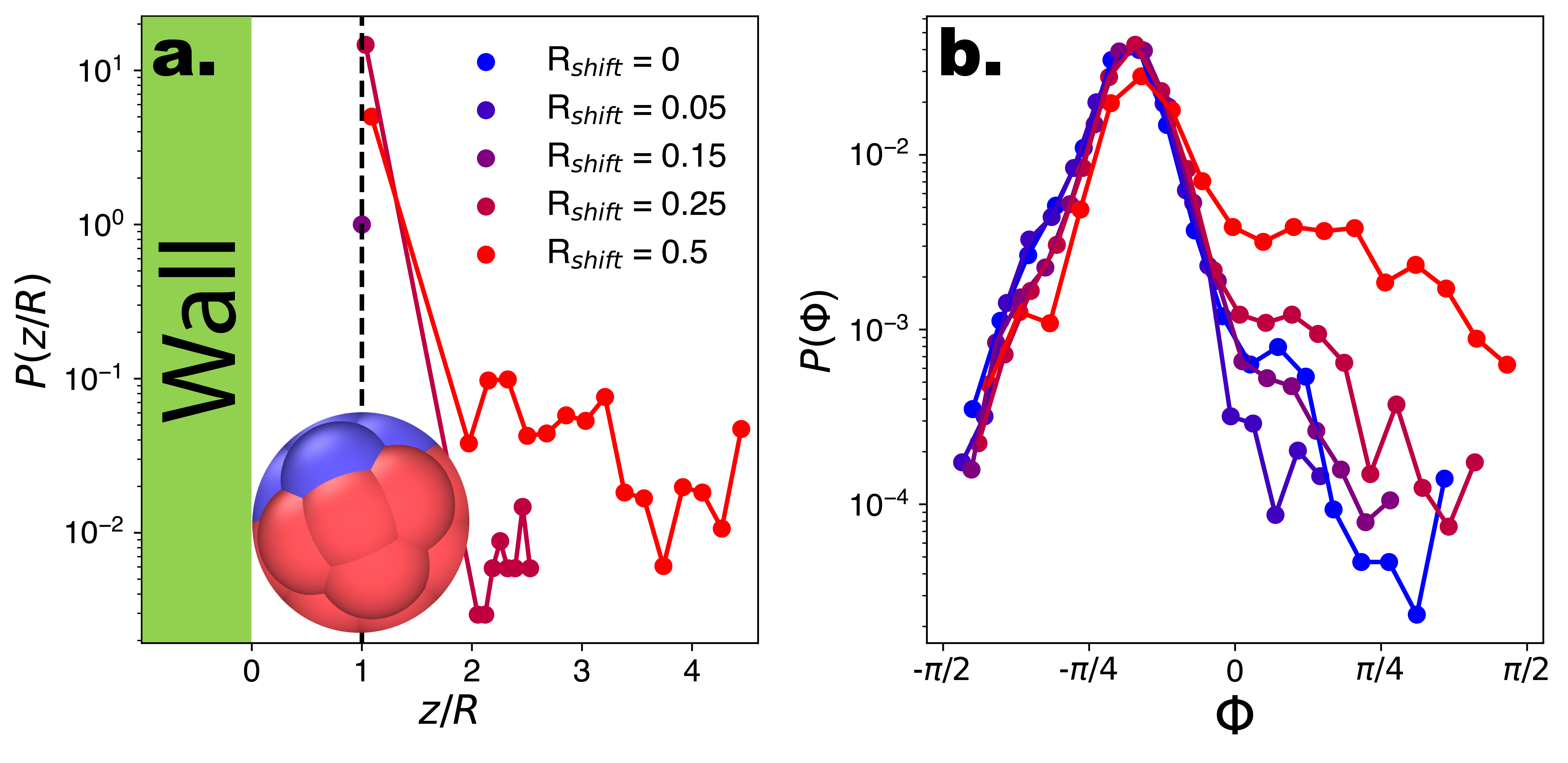}
  \caption{Probability distributions for CF microswimmers describing their out-of-plane motion $P(z/R)$ and orientation $P(\Phi)$, as a function of their introduced shape asymmetry $R_{shift}$, indicated by the colour code in the legend. a) Above a threshold ($R_{shift} > 0.25$), the microswimmers are able to leave the substrate and enter the bulk. We note that all particles with $R_{shift} \leq 0.15$ cannot leave the substrate and therefore have an overlayed point at unity. For visualisation, $z$ where $z/R < 2$ is set to $R$, and the presence of the wall at $z/R = 0$ is indicated. The particle depicted at the wall has an orientation $\Phi \sim -\pi/8$.  b) The ability to leave the substrate is reflected in the emergence of a shoulder in $P(\Phi)$ for $\Phi > 0$, as $V_{swim} > V_{gravity}$ and therefore the microswimmer can overcome its sedimentation velocity and enter the bulk.}
  \label{fig:Fig4}
\end{figure}

After establishing the effect of shape asymmetry on microswimmer motion in the absence of external fields and confining boundaries, we then introduce gravity and a substrate to simulate the experimental conditions in \cite{Bailey2021a} using $m_{asymm} = 1.25$. We define $V_{swim}:V_{sediment} \sim 1.5$ to prevent the microswimmers from escaping too far into the bulk. Under these conditions, microswimmers with no or low shape asymmetry are unable to leave the bottom substrate due to the joint forces of gravity and hydrodynamic attraction (see Figure \ref{fig:Fig4} a.). However, beyond a threshold value of $R_{shift} > 0.25$, the microswimmers display 3-D motion and begin to loop into the bulk, demonstrated by the emerging tail in $P(z/R)$. This behaviour is mirrored in the shoulder of the probability distribution of $\Phi, P(\Phi)$ (Figure \ref{fig:Fig4} b.) - the angle that the internal orientation of the particle makes with the substrate (see Figure \ref{fig:Fig_Schematic} d.) - as the positive values indicate that the particle points upwards and away into the bulk, thus overcoming gravity. We thus propose that the angular velocity $\omega$, introduced by the shape asymmetry of the microswimmers, drives its internal swimming orientation away from the substrate, competing with the effects of gravity and hydrodynamic wall interactions (as well as thermal fluctuations). Above a certain threshold shape asymmetry - governed by the dynamics of the system - the microswimmer is able to escape the substrate and move out-of-plane. 

In summary, we demonstrated that shape asymmetry is an important design parameter to control the dynamics of chemical microswimmers, in particular to promote 3-D motion. Specifically, the presence of these introduced asperities explains the unconventional swimming patterns observed in some experimental systems \cite{Bailey2021a,Bailey2023}.

\section{Conclusions}
By extending the DPD numerical framework described in \cite{Barriuso2022} to include the effects of mass and shape asymmetries, we have demonstrated that hydrodynamic interactions, gravity, and thermal fluctuations are sufficient to capture the dynamics of a range of experimental chemically active colloidal systems \cite{Niggel2023,Carrasco2023a,Bailey2021a,Bailey2023}. Promisingly, the use of DPD particles to build the overall structure of the microswimmer enables us to access to a range of more complex shapes than those described here, whose formulation would otherwise be either intractable or highly challenging when using other numerical modelling frameworks, or explicitly including chemical fields. We believe that our numerical simulations could be used in the design of new chemical colloidal swimmers, e.g. via sequential-capillarity-particle-assembly (sCAPA) \cite{Ni2017} or two-photon polymerization direct laser writing (2PP-DLW) \cite{Doherty_2020}, with the goal of realising chemical microswimmers capable of 3-D motion. Furthermore, this modular approach to generating complex structures enables the study of microswimmer physics considering arbitrary geometries, e.g. swimming above rough surfaces \cite{Carrasco2023a}, informing experiments into the interactions of active materials with confining boundaries commonplace in applied settings. To conclude, we foresee DPD based approaches to microswimmer modelling to provide opportunities in the development and application of chemically active colloids.

\medskip
\textit{Acknowledgements} - M.R.B. acknowledges financial support from the ETH Zurich Doc.Mobility Fellowship scheme. I.P. acknowledges support from Ministerio de Ciencia MCIN/AEI/FEDER for financial support under grant agreement PID2021-126570NB-100 AEI/FEDER-EU, and from Generalitat de Catalunya under Program Icrea Academia and project 2021SGR-673. C.V. acknowledges financial support from MINECO under grant agreements EUR2021-122001, PID2019-105343GB-I00, IHRC22/00002, and PID2022-140407NB-C21 
 
\medskip
\textit{Author Contribution Statement} - Author contributions are defined based on the CRediT (Contributor Roles Taxonomy). Conceptualization: M.R.B., C.M.B.G., L.I., C.V. Discussions: M.R.B., C.M.B.G., J.M.R., V.N., V.C., I.B., I.P., L.I., C.V. Formal Analysis: M.R.B., C.M.B.G. Funding acquisition: M.R.B., L.I. Investigation: M.R.B. Methodology: M.R.B., C.M.B.G., J.M.R. Software: M.R.B., C.M.B.G., J.M.R. Supervision: L.I., C.V. Validation: M.R.B. Visualization: M.R.B., C.M.B.G., L.I. Writing - original draft: M.R.B., C.M.B.G., L.I. Writing - review and editing: M.R.B., C.M.B.G., J.M.R., V.N., I.B., I.P., L.I., C.V.     

\newpage


\begin{center}
\section*{\centering{The Supporting Information includes:}}
\end{center}

\renewcommand{\figurename}{Figure S}
\renewcommand{\tablename}{Table S}
\setcounter{figure}{0}    
\setcounter{table}{0}   


\begin{itemize}
    \item Fig. S1: Sedimentation of passive colloids under gravity
    \item Fig. S2: Trajectory of a microswimmer which leaves the plane
    \item Fig. S3: Velocity Distributions binned by different definitions of swimming direction
    \item Figure S4: Effect of thermal noise on the dynamics of a microswimmer without shape asymmetry
    \item Figure S5: Rendering of different shape asymmetries (1 Edge, 1 Centre, 2 Edges)
    \item Figure S6: Role of shape asymmetry on the dynamics of a microswimmer (1 Edge case)
    \item Figure S7: Role of shape asymmetry on the dynamics of a microswimmer (1 Centre case)
    \item Figure S8: Role of shape asymmetry on the dynamics of a microswimmer (2 Edges case)
\end{itemize}

\newpage

\begin{figure}[h]
\centering
   \includegraphics[width=\linewidth]{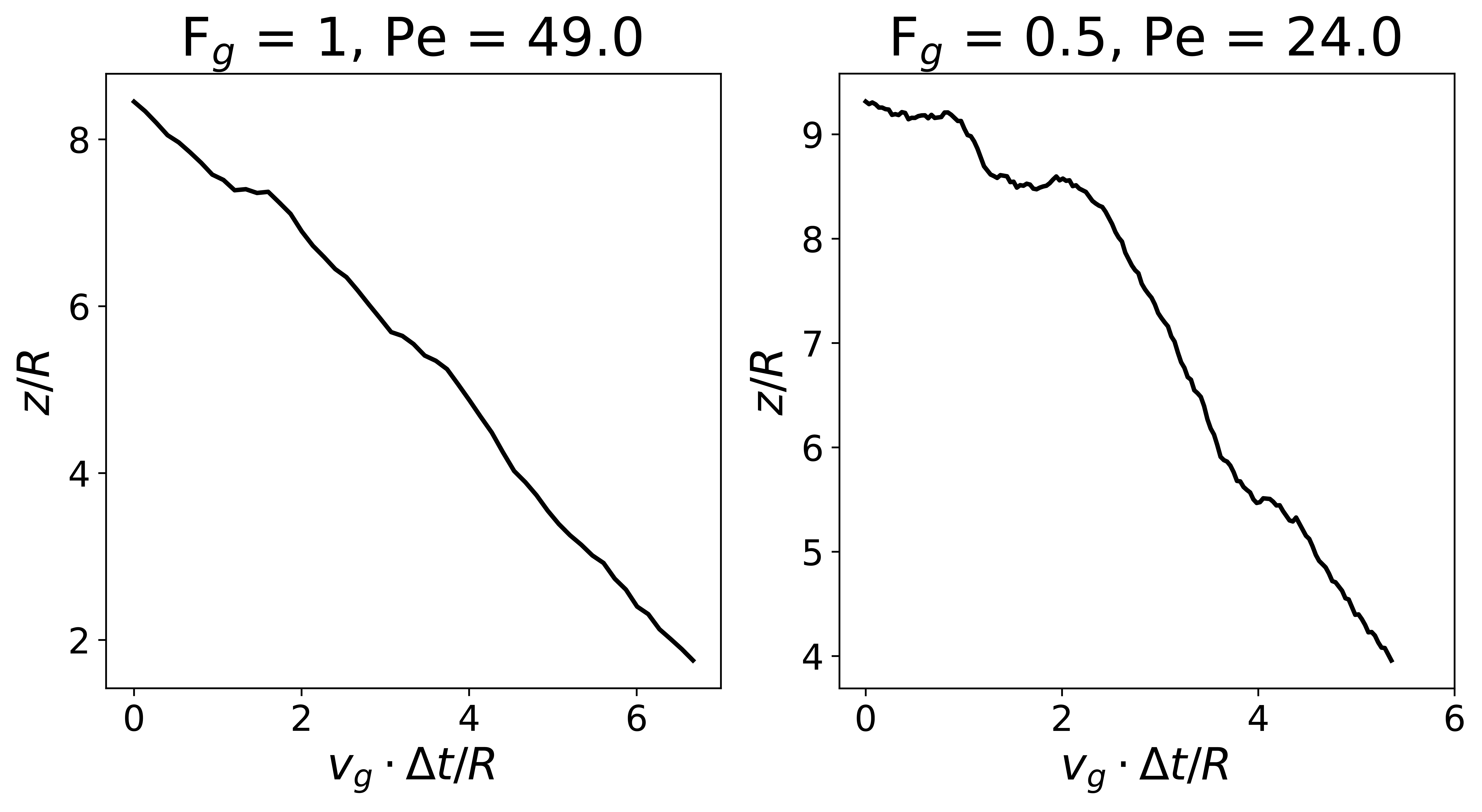} 
  \caption{Verification of the solvent parameters used to ensure a low Re regime. The gravitational force $F_g$ applied to a sedimenting colloid is halved (from $F_g = 1$ to $F_g = 0.5$, left panel and right panel respectively), with the corresponding reduction in sedimentation velocity and thus Pèclet number ($Pe = v_g*R/D_T$, where $v_g$ is the velocity due to gravity, and $R, D_T$ are constant properties of the colloid) - as expected by Stokes law. We note that the forces involved are relatively low when F\textsubscript{g} = 0.5, leading to increasingly important role of Brownian fluctuations. As the gravitational force is increased further, dz/dt will become increasingly linear (as Pe increases).}
   \label{suppfig:dzdt}
\end{figure}

\begin{figure}[h]
\centering
   \includegraphics[width=\linewidth]{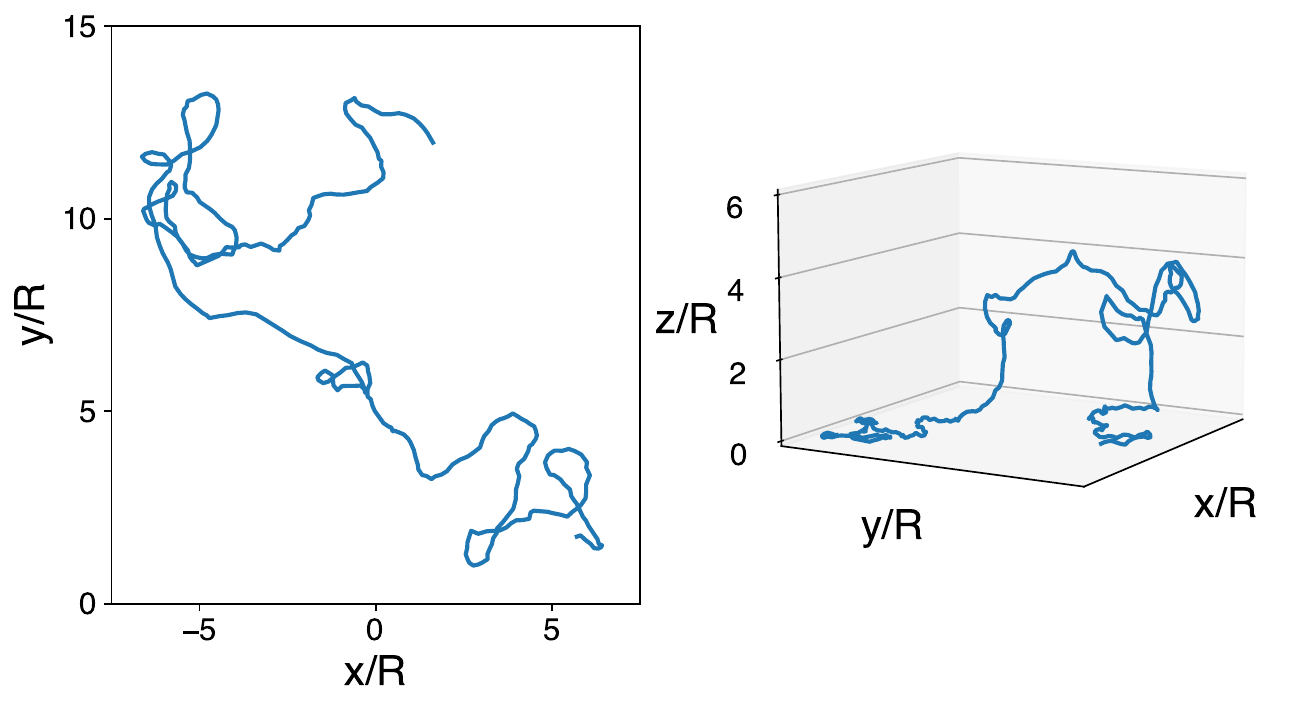} 
  \caption{An example trajectory of the 3-D motion of a microswimmer (right panel), with its 2-D projection onto the XY plane (left panel). We note that particle motion is characterised by long periods of motion in the 2-D plane, interdispersed with out-of-plane (z) loops.}
   \label{suppfig:dzdt}
\end{figure}

\begin{figure}[h]
\centering
   \includegraphics[width=\linewidth]{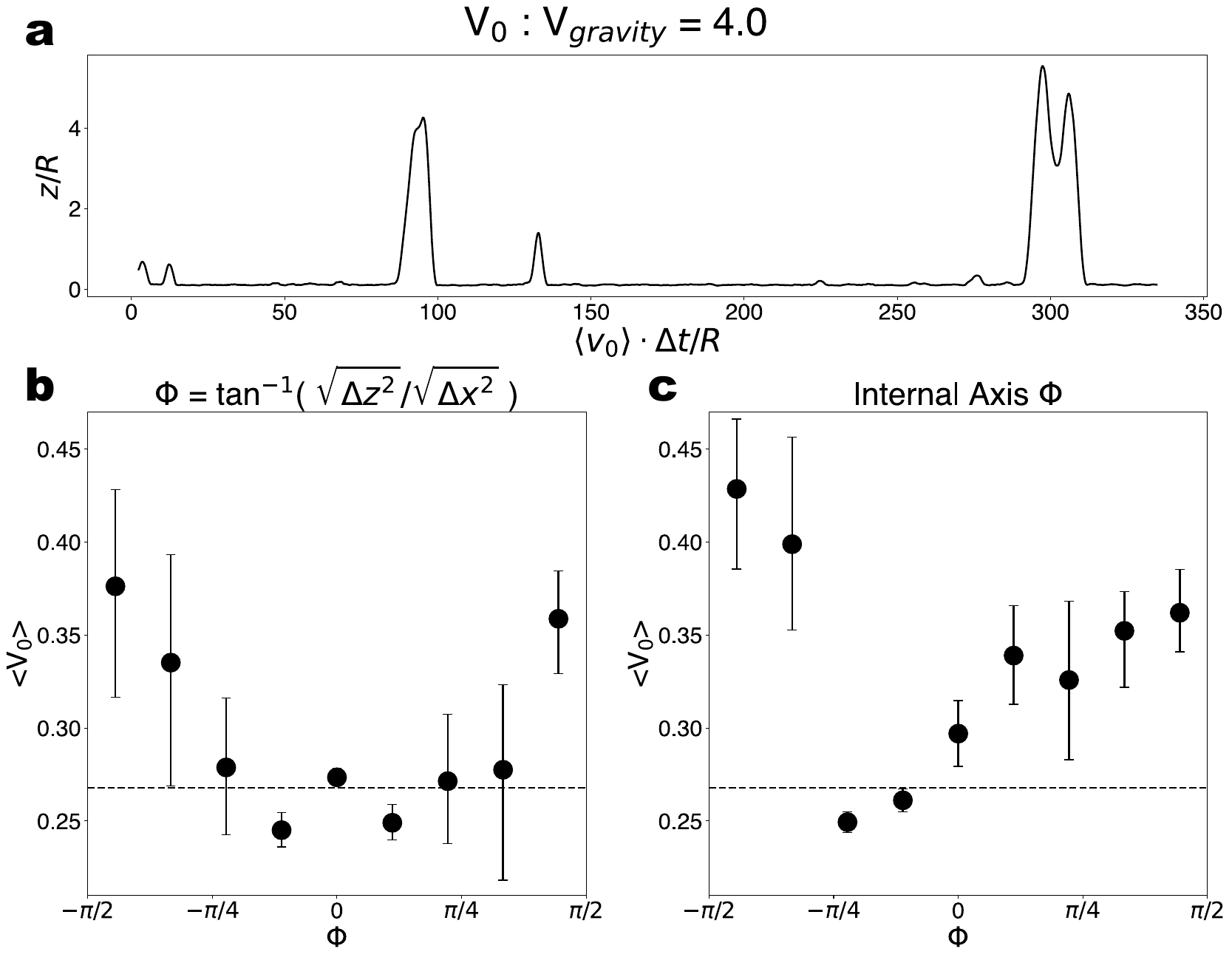} 
  \caption{a) The out-of-plane trajectory of a simulated CB microswimmer (z(t)). b) Using the analysis proposed in \cite{Bailey2021a,Bailey2023}, i.e. binning the instantaneous displacements of the microswimmer by the zenith angle calculated from those displacements, results in the same data structure as that found experimentally \cite{Bailey2021a,Bailey2023}. This suggests that the “orientation-dependent velocity" previously reported is in fact an inherent feature of the analytical approach used, rather than arising from e.g. the microswimmers' photo-responsiveness. c) Analysing the data using the internal orientation axis (known in simulations), we recover a more explainable structure. Specifically, when the particle swims down from the bulk towards the substrate ($-\pi/2 < \Phi < -\pi/4$), it obtains the expected boost in velocity due to gravity. For $-\pi/4 < \Phi < 0$, the velocities are significantly lower, as this corresponds to the region where the particle is at the substrate and facing towards it due to hydrodynamic attraction (see Figure 2 h.). The particle speed here is lower as it only consists of the projection of the velocity onto the substrate. We note that this corresponds to the largest part of the trajectory, significantly reducing the overall average velocity (dashed line). As the particle leaves the substrate ($0 < \Phi$), the particle swims faster than at the substrate, but as it swims against gravity is not as fast as for $-\pi/2 < \Phi < -\pi/4$. These results suggest that coarse-graining the effect of phoretic and light fields to hydrodynamic interactions and gravity is indeed sufficient to capture the physics of various experimental chemical microswimmer systems.}
   \label{suppfig:veldistbn} 
\end{figure}

\begin{figure}[h]
\centering
   \includegraphics[width=\linewidth]{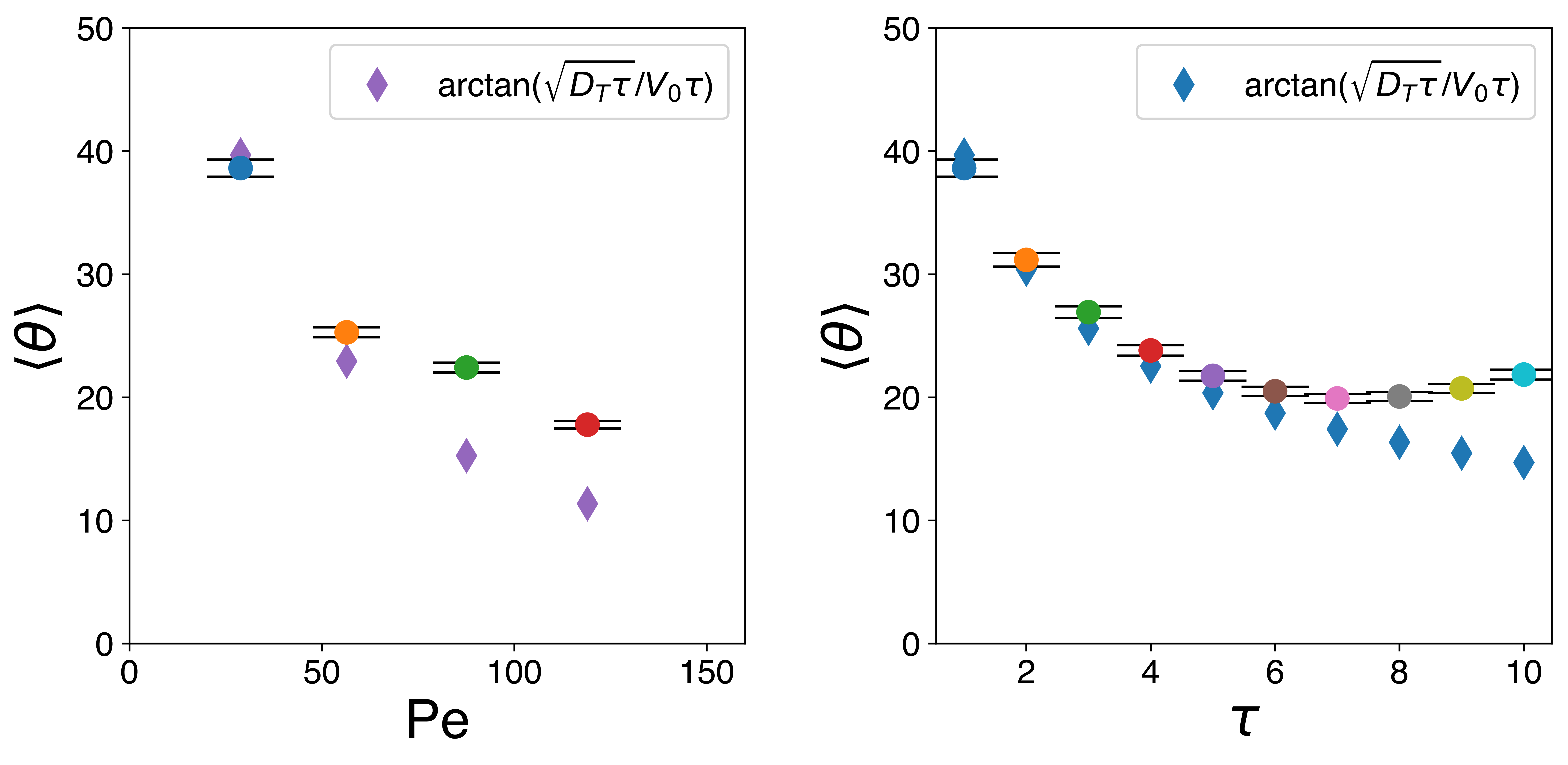} 
  \caption{Investigation of the angle $\theta$ between the swimming direction and internal orientation (defined by the body axis) of our microswimmers in bulk and in the absence of any shape asymmetry ($R_{shift} = 0$). Left panel: The change in $\theta$ as the swimming velocity of the microswimmer is increased (represented via $Pe = v_p\cdot R/D_T$. Right panel: The change in $\theta$ as the lag-time $\tau$ over which the subsequent displacements are evaluated (see main text for discussion of definitions). In both cases, the diamonds represent the values estimated if assuming translational diffusion occurs only perpendicular to the internal swimming axis. As no gravitational force nor substrate is present, one may expect that the internal orientation and swimming direction of a weak pusher squirmer would align perfectly. However, from our analysis we find that this is not the case. We propose that thermal fluctuations, namely rotational and translational diffusion of the particle, are the key contributing factors to this observation. On the left, we see that at lower Pe numbers, the significant divergence between the swimming direction and internal orientation can be completely explained by the expected translational diffusion of the particle. We find that as the Pe is increased, both the observed and theoretical $\theta$ between the two vectors follow the same trend, in that the value of $\langle\theta\rangle$ decreases. We also see on the right that by increasing lag time over which the displacements are evaluated, the observed and expected $\langle\theta\rangle$ values also become smaller. This points to the role of the microswimmer's diffusivity in defining its motion, and thus the angle between the swimming direction and the body axis in the case of $R_{shift} = 0$ (no shape asymmetry). Due to finite time differencing, we also expect the rotational diffusivity of the squirmer to play a further role in the difference between the particles internal orientation and swimming displacement in bulk. For example, we take the average of the particle's internal orientation between two time steps, which will introduce a component of noise due to rotational diffusion between those steps. We also expect these effects to contribute to the saturation of the effect of shape asymmetry at $\theta \sim$ 35$^{\circ}$. }
   \label{suppfig:thermalNoise} 
\end{figure}

\begin{figure}[h]
\centering
   \includegraphics[width=\linewidth]{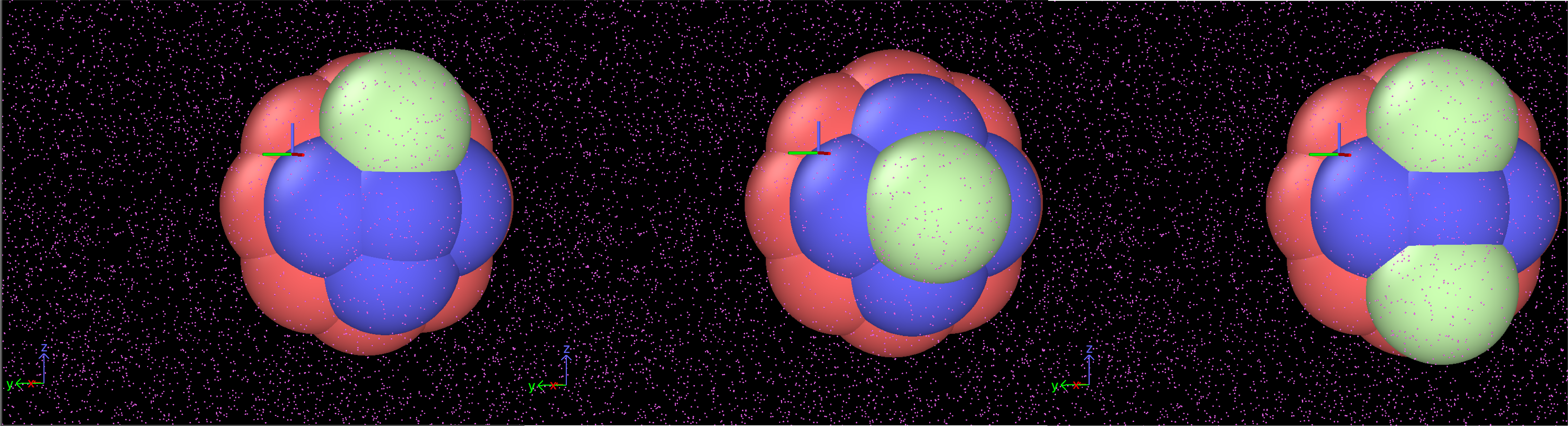} 
  \caption{Overview of the different types of shape asymmetry introduced via shape particles $P_s$ with $R_{shift} = 0.5$ (light green) along the axes of different “cap" particles (blue) for CF microswimmers. Left panel: “1 Edge" case, where $P_s$ is introduced at a $45^\circ$ angle to the internal body axis (central blue particle). Middle panel: “1 Centre" case, where $P_s$ is introduced along the internal body axis. Right panel: “2 Edge" case, where the two $P_s$ particles are introduced at $45^\circ$ and $- 45^\circ$ to the internal body axis (central blue particle). In all cases, the swimming direction is with the blue cap pointing forwards. The red particles are the remaining “filler" particles constituting the rest of the microswimmer body. Solvent particles are reduced for scale and represented in purple. Graphics were generated using the visualisation software Ovito \cite{Stukowski2010}}
   \label{suppfig:shapeasymms} 
\end{figure}

\begin{figure}[h]
\centering
   \includegraphics[width=\linewidth]{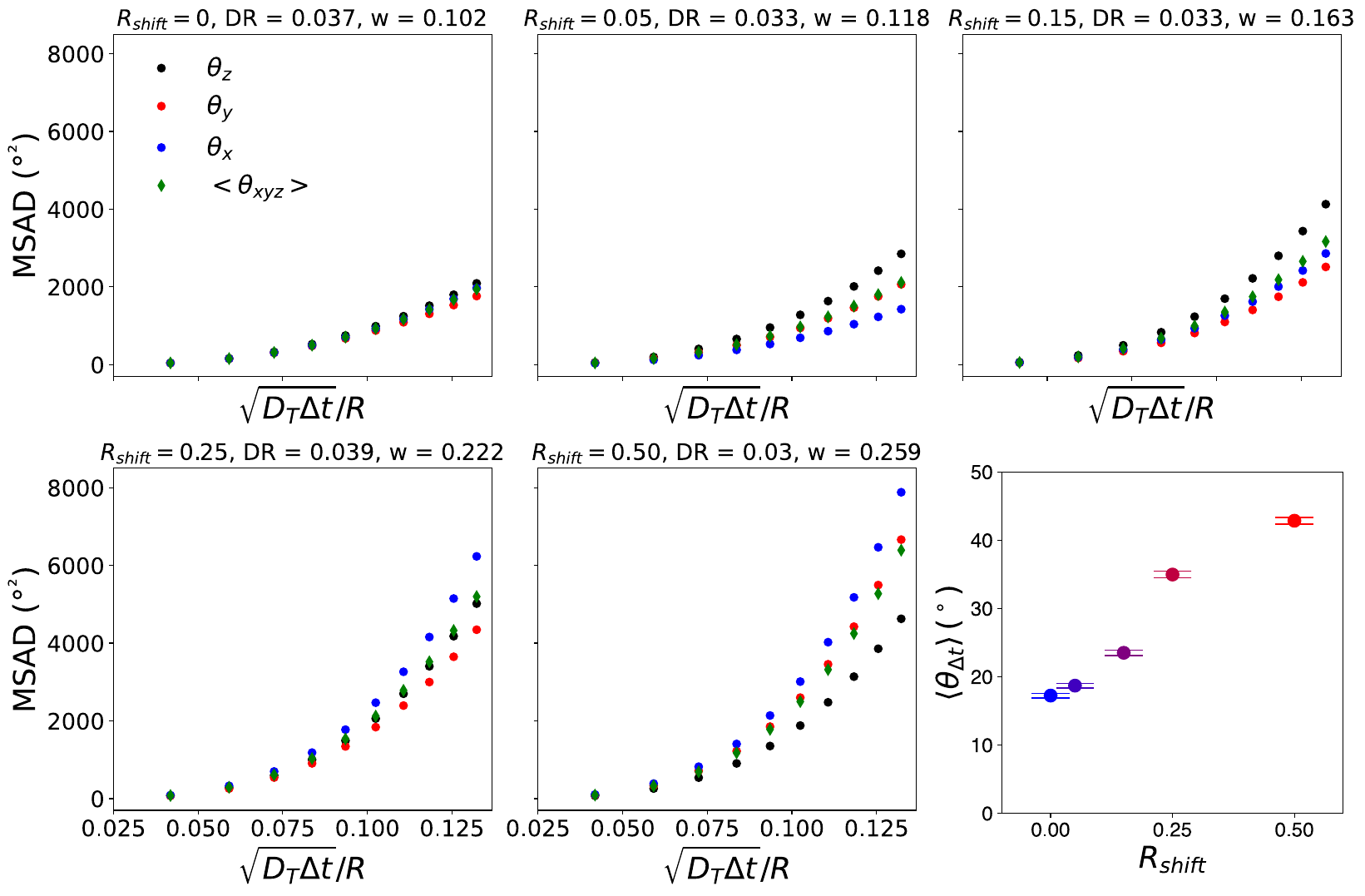} 
  \caption{MSADs for the simulated bulk microswimmers for increasing values of $R_{shift}$ for the “1 Edge" case, as described in Figure S\ref{suppfig:shapeasymms}. The calculated angles between the swimming direction and internal orientation of bulk microswimmers, also discussed in Figure 4 of the main text, are shown in the bottom right panel. Error bars indicate the standard error of the mean from 1001 frames (sub-sampled from 50000 simulation steps).}
   \label{suppfig:1Edge_MSAD} 
\end{figure}

\begin{figure}[h]
\centering
   \includegraphics[width=\linewidth]{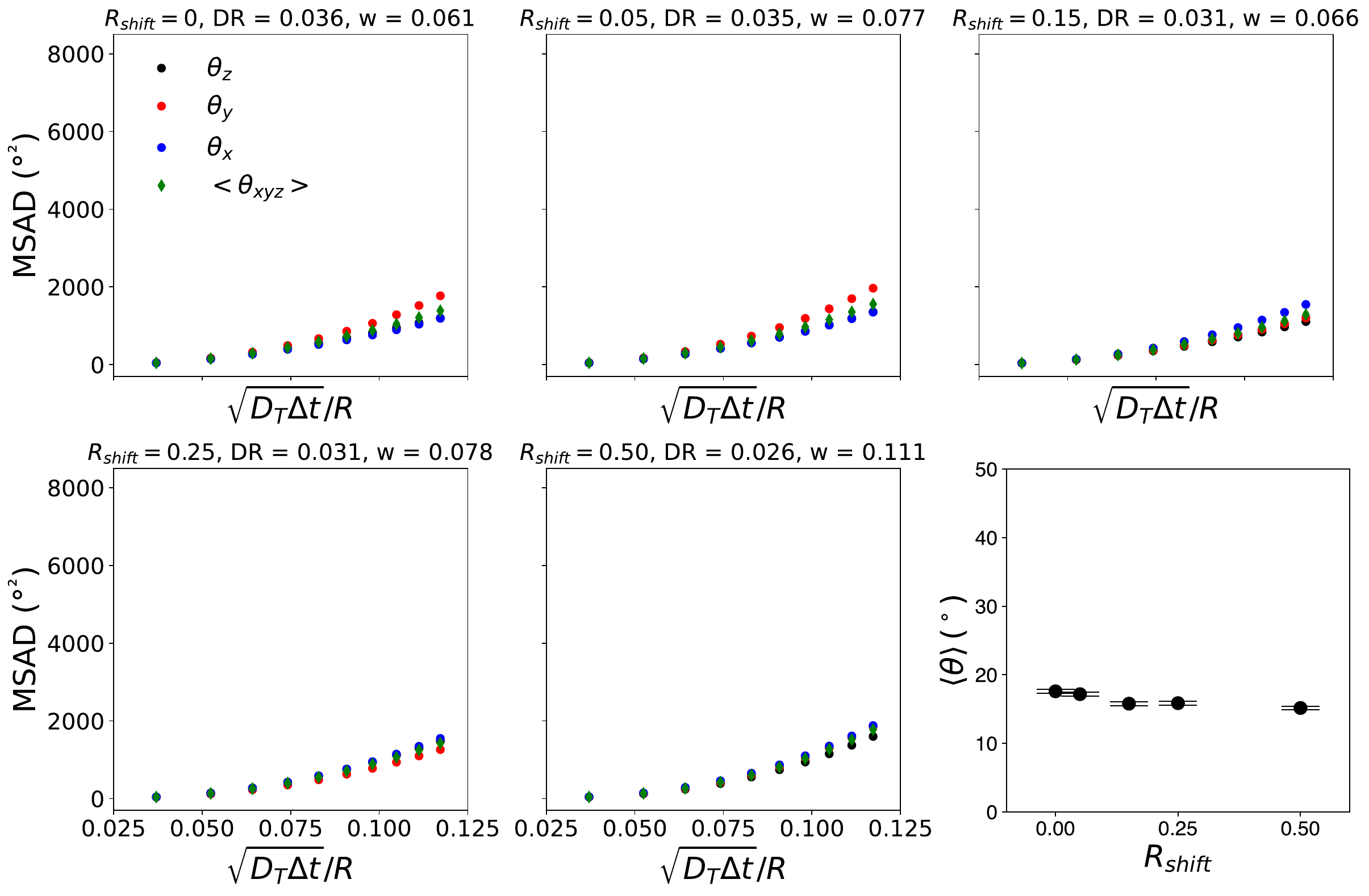} 
  \caption{MSADs for the simulated bulk microswimmers for increasing values of $R_{shift}$ for the “1 Centre" case, as described in Figure S\ref{suppfig:shapeasymms}. The calculated angles between the swimming direction and internal orientation of bulk microswimmers are shown in the bottom right panel. We note that the reduced asymmetry introduced by a $P_{shape}$ particle along the internal microswimmer orientation has little effect on its dynamics, contrasting to the “1 Edge" case. Error bars indicate the standard error of the mean from 1001 frames (sub-sampled from 50000 simulation steps).}
   \label{suppfig:1Centre_MSAD} 
\end{figure}

\begin{figure}[h]
\centering
   \includegraphics[width=\linewidth]{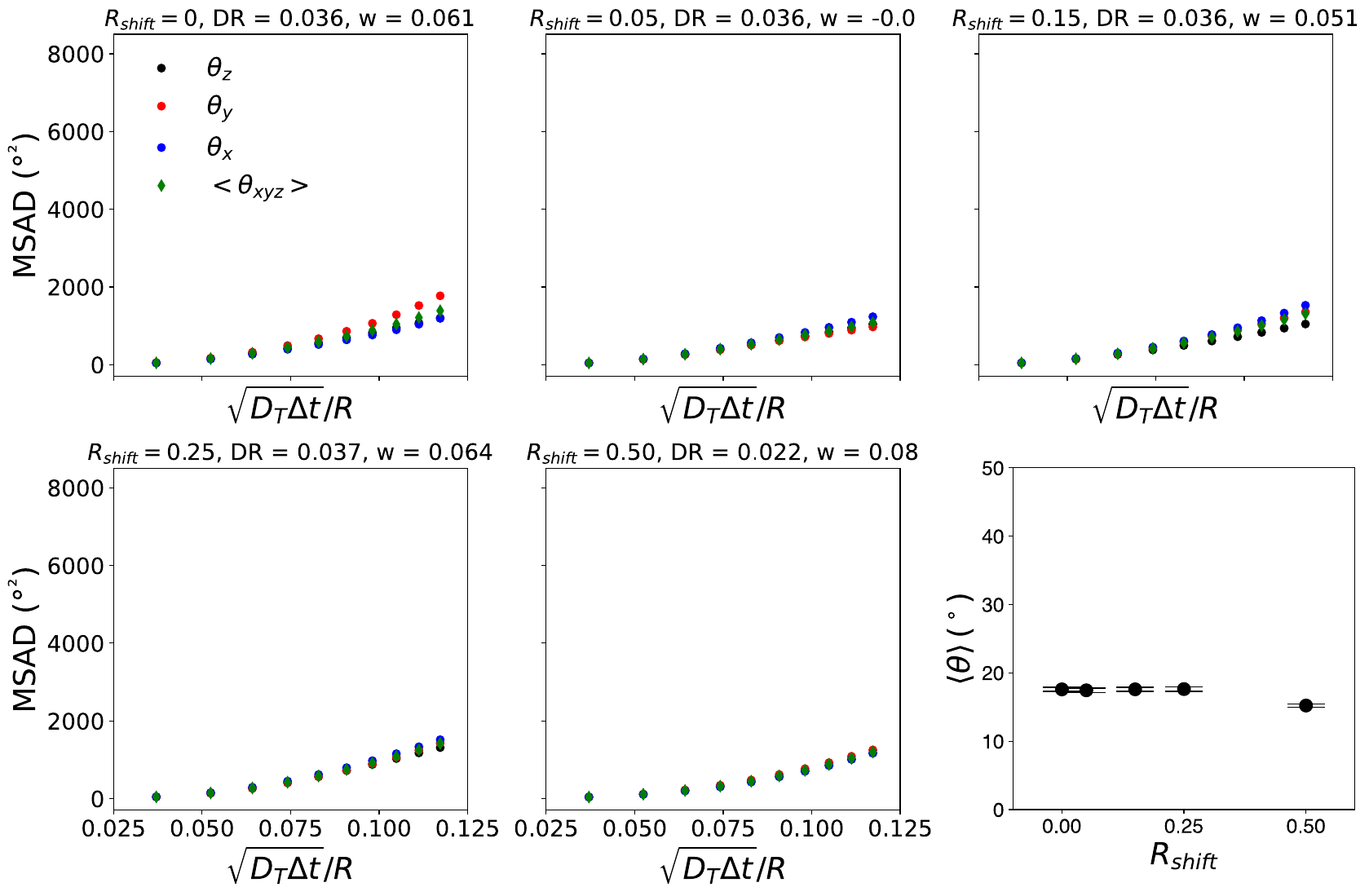} 
  \caption{MSADs for the simulated bulk microswimmers for increasing values of $R_{shift}$ for the “2 Edge" case, as described in Figure S\ref{suppfig:shapeasymms}. The calculated angles between the swimming direction and internal orientation of bulk microswimmers are shown in the bottom right panel. We note that the reduced asymmetry introduced by 2 $P_{shape}$ particles on the edges of the microswimmer, resulting in mirror symmetry along its internal orientation axis, has little effect on the swimming dynamics, contrasting to the “1 Edge" case. This is despite the introduction of 2 $P_{shape}$ particles on positions each corresponding to the “1 Edge" case (opposite each other). This underlines the importance of the overall shape asymmetry of the microswimmer to its physics. Error bars indicate the standard error of the mean from 1001 frames (sub-sampled from 50000 simulation steps).}
   \label{suppfig:2Edge_MSAD} 
\end{figure}



\end{document}